\documentclass[journal, twocolumn,dvipdfmx]{IEEEtran}

\usepackage{setspace}
\usepackage{graphicx}
\usepackage{subfigure}
\usepackage{tabularx}
\usepackage{arydshln}
\usepackage{umoline}
\usepackage[dvips]{color}
\usepackage{picinpar}
\usepackage{mathrsfs}
\usepackage{latexsym}
\usepackage{amssymb}
\usepackage{pifont}
\usepackage{amsfonts}
\usepackage{amsmath}
\usepackage{color}
\usepackage{fancyhdr}
\usepackage{float}
\usepackage{cases}
\usepackage{bm}
\usepackage{arraycols}
\usepackage{bbding}

\usepackage{epsfig}
\usepackage{graphics}
\usepackage[Lenny]{fncychap}
\usepackage{citesort}
\usepackage{subsections}
\usepackage{rotfloat}
\usepackage{dropping}
\usepackage{enumitem}
\usepackage[normalem]{ulem}

\def\c{\mathbf{c}}

\def\v{\mathbf{v}}

\ifCLASSINFOpdf
\else
\fi

\begin{document}
%
%
\title{
Precoded Faster-than-Nyquist Signaling Using Optimal Power Allocation for OTFS}
\author{
Zekun~Hong,
Shinya~Sugiura,~\IEEEmembership{Senior~Member,~IEEE},
Chao~Xu,~\IEEEmembership{Senior~Member,~IEEE},
and Lajos~Hanzo,~\IEEEmembership{Life~Fellow,~IEEE}
\thanks{Copyright $\copyright$ 2024 IEEE. Personal use of this material is permitted. However, permission to use this material for any other purposes must be obtained from the IEEE by sending a request to pubs-permissions@ieee.org.}
\thanks{Z. Hong and S. Sugiura are with the Institute of Industrial Science, The University of Tokyo, 153-8505, Japan (e-mail: sugiura@iis.u-tokyo.ac.jp). C. Xu and L. Hanzo are with the School of Electronics and Computer Science, University of Southampton, SO17 1BJ, UK. (\textit{Corresponding author: Shinya Sugiura}.)}
\thanks{The work of Z. Hong was supported in part by JST SPRING (Grant JPMJSP2108). The work of S. Sugiura was supported in part by JST FOREST (Grant JPMJFR2127), in part by JST ASPIRE (Grant JPMJAP2345), and in part by JSPS KAKENHI (Grants 22H01481, 23K22752).
L. Hanzo would like to acknowledge the financial support of the Engineering and Physical Sciences Research Council (EPSRC) projects under grant EP/Y037243/1, EP/W016605/1, EP/X01228X/1, EP/Y026721/1, EP/W032635/1 and EP/X04047X/1 as well as of the European Research Council's Advanced Fellow Grant QuantCom (Grant No. 789028).}
\vspace{-8mm}
}
\markboth{Preprint (Accepted Version) for publication in IEEE Wireless Communications Letters (DOI: 10.1109/LWC.2024.3491777)}
{Shell \MakeLowercase{\textit{et al.}}: Bare Demo of IEEEtran.cls for Journals}
\maketitle
\begin{abstract}
A precoded orthogonal time frequency space (OTFS) modulation scheme relying on faster-than-Nyquist (FTN) transmission over doubly selective fading channels is {proposed}, which enhances the spectral efficiency and improves the Doppler resilience. We derive the input-output relationship of the FTN signaling in the delay-Doppler domain. Eigenvalue decomposition (EVD) is used for eliminating both the effects of inter-symbol interference and correlated additive noise encountered in the delay-Doppler domain to enable efficient symbol-by-symbol demodulation. Furthermore, the power allocation coefficients of individual frames are optimized for maximizing the mutual information under the constraint of the {derived} total transmit power. Our performance results demonstrate that the proposed FTN-based OTFS scheme can enhance the information rate while achieving a comparable BER performance to that of its conventional Nyquist-based OTFS counterpart that employs the same root-raised-cosine shaping filter.
\end{abstract}
\begin{IEEEkeywords}
faster-than-Nyquist signaling, OTFS, mutual information, information rate, precoding, doubly selective fading.
\end{IEEEkeywords}

\IEEEpeerreviewmaketitle
\vspace{-7mm}
\section{Introduction}
\IEEEPARstart{T}{raditional}
wireless technologies, such as orthogonal frequency-division multiplexing (OFDM), typically struggle to cope with high Doppler shifts in high-mobility scenarios.
In order to overcome this limitation, orthogonal time frequency space (OTFS) modulation~\cite{hadani2017orthogonal} was proposed, which modulates data symbols in the delay-Doppler (DD) domain and allows a sparse and quasi-static channel representation.
Several studies~\cite{farhang2017low,xu2022otfs,10453468} considered the OFDM-based OTFS (OFDM-OTFS) architecture, where a cyclic prefix (CP) is inserted in each frame, while employing an ideal rectangular pulse shaping filter that satisfies bi-orthogonality in the time and frequency domains~\cite{hadani2017orthogonal}.
By contrast, the input-output relationship of OTFS using a non-rectangular pulse shaping filter was derived in \cite{raviteja2018practical}.

The symbol interval of faster-than-Nyquist (FTN) signaling is designed to be lower than that given by the Nyquist criterion. Hence, FTN signaling allows us to boost the transmission rate without increasing the bandwidth requirement, albeit at the cost of producing inevitable inter-symbol interference (ISI) at the receiver, as detailed in~\cite{ishihara2021evolution}.
Several precoded FTN schemes were proposed for mitigating the impairments encountered~\cite{kim2016properties,ishihara2019svd,ishihara2021eigendecomposition,ishihara2022reduced}.
In \cite{kim2016properties}, precoding-aided FTN signaling based on the eigenvalue decomposition (EVD) of an FTN-specific ISI matrix was proposed. In \cite{ishihara2019svd}, EVD-precoded FTN signaling relying on optimal power allocation (PA) was developed to improve the information rate.
In \cite{ishihara2021eigendecomposition}, the EVD-precoded FTN signaling scheme of \cite{ishihara2019svd} was extended to that applicable to generalized frequency-selective fading channels.
In \cite{ishihara2022reduced}, fast Fourier transform (FFT)-spread multi-carrier FTN (MFTN) signaling was proposed by relying on the circulant approximation of the FTN-specific ISI matrix and the noise covariance matrix.
{Additionally, several studies discussed data detection~\cite{wu2016frequency,ishihara2023differential,zhou2022precoded} and channel estimation~\cite{keykhosravi2023pilot,10552123} for FTN signaling transmission over doubly selective fading channels to improve the robustness to the Doppler shift.}
{However, all previous OTFS studies assumed that the transmitted signal obeys the time-orthogonal Nyquist criterion, similar to conventional signaling.
Moreover, none of the above FTN signaling schemes have been designed to mitigate doubly selective fading in the DD domain.}

Against the above background, the novel contributions of this paper are as follows.
\begin{itemize}
\item
For the first time in literature, we propose an EVD-precoded OTFS-based FTN (OTFS-FTN) architecture for doubly selective fading channels. We will demonstrate that the proposed OTFS-FTN scheme exhibits robustness to high Doppler shifts, while enhancing the spectral efficiency as a benefit of the reduced FTN symbol interval. We derive the input-output relationship between the transmitted and received symbols of the proposed OTFS-FTN scheme employing a root-raised-cosine (RRC) pulse shaping filter in the DD domain.
Furthermore, both the effects of FTN-specific ISI and of the correlated noise are eliminated with the aid of EVD-based diagonalization in the DD domain, hence allowing efficient symbol-by-symbol demodulation at the receiver.
\item
As an additional contribution, we derive the mutual information (MI) characterizing the proposed scheme and then design the PA to maximize the MI for transmission over doubly selective fading channels.
\item Our simulation results will demonstrate that the proposed OTFS-FTN scheme achieves a higher information rate and better BER performance than its conventional time-orthogonal signaling counterpart using the same RRC shaping filter.
\end{itemize}

\section{System Model}
\label{model}
\subsection{Transmit Signal}
{Each transmission block of the proposed OTFS-FTN scheme has $M$ subcarriers and $N$ time slots.}
The information symbols $\mathbf{x}=[x_{0}, \cdots, x_{MN-1}]^T \in {\mathbb{C}^{MN}}$ are precoded by a matrix $\mathbf{P}\in {\mathbb{C}^{MN \times MN}}$, yielding {precoded symbols} $\mathbf{x}_\mathrm{p}=\mathbf{P}\mathbf{x}$.
Note that average symbol energy is defined as $\mathbb{E}\left[\left|x_{n}\right|^2\right]=\sigma_{\mathrm{x}}^2 \ (n=0,\cdots, MN-1)$.

Assuming a subcarrier spacing of $\Delta f$ and that $T=1/\Delta f$, the frame interval and bandwidth are given by $NT$ and $M \Delta f$, respectively.
Furthermore, the FTN-specific sampling interval is represented by $T_{\mathrm{f}}={T}/{M}=\alpha T_{0}$, where $\alpha$ is the FTN packing ratio and $T_0=1/(2W)$ is the symbol interval defined by the Nyquist criterion, {which corresponds to conventional OTFS modulation.} {Furthermore, $2W$ represents the bandwidth of an ideal rectangular shaping filter.}
Similar to typical OFDM-OTFS schemes~\cite{xu2022otfs,10453468}, in order to convert the modulated symbols in the DD domain to the transmitted symbols in the time-frequency (TF) domain, the inverse symplectic fast Fourier transform (ISFFT) is carried out, which corresponds to the $M$-point FFT for the columns and the $N$-point IFFT for the rows in {precoded symbols} $\mathbf{X}_\mathrm{p} \in {\mathbb{C}^{M \times N}}$.
Then, an $M$-point IFFT is utilized for generating the time-domain (TD) signal, which is expressed by
\begin{IEEEeqnarray}{rCL}
\mathbf{S}=\mathbf{F}_M^{H}\left(\mathbf{F}_M \mathbf{X}_\mathrm{p} \mathbf{F}_N^{H}\right),\label{eq:S}
\end{IEEEeqnarray}
where we consider the relationship of $\mathbf{x}_\mathrm{p}=\mathrm{vec}({\mathbf{X}_\mathrm{p}})$, {where $\mathrm{vec}(\cdot)$ represents the column-wise vectorization.}
Moreover, $\mathbf{F}_M \in \mathbb{C}^{M \times M}$ and $\mathbf{F}_N\in \mathbb{C}^{N \times N}$ represent the $M$-point and $N$-point normalized DFT matrices, respectively.
More specifically, the $k$th-row and $m$th-column entry of $\mathbf{F}_N$ is given by $\frac{1}{\sqrt{N}} e^{-2 \pi j (k-1) (m-1) / N}$.

Then, the column-wise vectorization of $\mathbf{S}$ is represented by
\begin{IEEEeqnarray}{rCL}
\mathbf{s}=\mathrm{vec}({\mathbf{S}})=\left(\mathbf{F}_N^{H} \otimes \mathbf{I}_{M}\right) \mathbf{x}_\mathrm{p},\label{eq:s}
\end{IEEEeqnarray}
where $\otimes$ is the Kronecker product, and $\mathbf{I}_{M} \in \mathbb{R}^{M \times M}$ is the identity matrix.

For {an} RRC pulse shaping filter $h_{\mathrm{tx}}(t)$ having the roll-off factor $\beta$, the baseband OTFS-FTN transmit signal is given by
\begin{IEEEeqnarray}{rCL}
s(t)= \sum_{n=0}^{MN-1} s_{n} h_{\mathrm{tx}}(t-nT_{\mathrm{f}}). \label{eq:s_t}
\end{IEEEeqnarray}

Similar to \cite{raviteja2018practical}, a cyclic prefix is inserted at the beginning of each OTFS-FTN frame.
Note that by adding a sufficiently long CP, the detrimental effects of inter-frame interference caused by FTN transmission can be ignored for the practical range of packing ratios, i.e., for $\alpha \geq1/(1+\beta)$~\cite{ishihara2022reduced,ishihara2023differential}.

\subsection{Channel Model}
The signal $r(t)$, which is received over the time-varying channel, is formulated as \cite{hadani2017orthogonal}:
\begin{IEEEeqnarray}{rCL}
r(t)=\iint h(\tau, \nu) s(t-\tau) e^{j 2 \pi \nu(t-\tau)} d \tau d \nu+w(t), \label{eq:rt}
\end{IEEEeqnarray}
where $\tau$ and $\nu$ denote the delay and Doppler shift, respectively.
Furthermore, $w(t)$ represents the complex-valued additive white Gaussian noise (AWGN), whose power spectral density is given by $\sigma_0^2$.

Owing to the sparsity of the DD channel, the channel response $h(\tau, \nu)$ can be expressed by
\begin{IEEEeqnarray}{rCL}
h(\tau, \nu)=\sum_{p=0}^{P-1}h_{p} \delta(\tau-\tau_p)\delta(\nu-\nu_p), \label{eq:h}
\end{IEEEeqnarray}
where $\delta(\cdot)$ is Dirac's delta function, and $P$ is the number of channel taps. Furthermore, $h_p$, $\tau_p$, and $\nu_p$ represent the complex-valued channel gain, the propagation delay, and the Doppler shift of the $p$th path.
More specifically, we have
\begin{IEEEeqnarray}{rCL}
\tau_p=\frac{l_p}{M \Delta f}, \quad \nu_p=\frac{k_p+\kappa_p}{N T},\label{eq:DDtaps}
\end{IEEEeqnarray}
where $l_p$ is the delay tap and $(k_p+\kappa_p)$ is the Doppler shift tap of the $p$th path, respectively.
Furthermore, $l_p$ and $k_p$ denote integers, while $\kappa_p$ represents the fractional part of the Doppler tap in the range of $-{1}/{2}<\kappa_{p} \leq {1}/{2}$.
Let us assume that the channel's maximum delay $\tau_{\max }$ and the maximum Doppler shift $\nu_{\max }$ satisfy $\tau_{\max } \leq (L-1) T / M$ and $\left|\nu_{\max }\right| \leq \Delta f / 2$, respectively, where $L$ is the CP length. Furthermore, some of the paths that share the same delay but have different Doppler shifts are separable only in the DD domain.
\subsection{Receiver Model}
According to \eqref{eq:s_t}, \eqref{eq:rt}, and \eqref{eq:h}, after the removal of the CP and following matched filtering by the pulse shaping filter $h_{\mathrm{rx}}^{*}(-t)$, the received signal $z(t)$ is given by
\begin{IEEEeqnarray}{rCL}
z(t)\!&=&\!\left[\!\sum_{p=0}^{P-1} h_p e^{j 2 \pi \frac{(k_p+\kappa_p)\left(t-l_p T_\mathrm{f}\right)}{M N T_{\mathrm{f}}}} s\left(t-l_p  T_{\mathrm{f}}\right)+w(t)\!\right]\!\star\! h_{\mathrm{rx}}^{*}(-t) \nonumber\\
&=&\sum_{p=0}^{P-1} \sum_{n=0}^{MN-1} h_p e^{j 2 \pi \frac{(k_p+\kappa_p)\left(t-l_p T_{\mathrm{f} }\right)}{M N T_{\mathrm{f}}}} s_n g\left(t-\left(n+l_p\right) T_{\mathrm{f}}\right)\nonumber\\ && +\eta(t), \label{eq:rt_m}
\end{IEEEeqnarray}
where $\star$ denotes the convolution operation, while we have $g(t) \triangleq h_{\mathrm{tx}}(t) \star h_{\mathrm{rx}}^{*}(-t)$ and $\eta(t) \triangleq w(t) \star h_{\mathrm{rx}}^{*}(-t)$.

By sampling $z(t)$ at $t=0,\cdots, (MN-1)T_{\mathrm{f}}$, we arrive at:
\begin{IEEEeqnarray}{rCL}
\mathbf{z}&=&[z_0,\cdots,z_{MN-1}]^T \ \in {\mathbb{C}^{MN}} \nonumber \\
&=&\mathbf{H}\mathbf{s}+\boldsymbol{\eta},
\end{IEEEeqnarray}
where $\boldsymbol{\eta}=[\eta(0), \eta(T_{\mathrm{f}}) \cdots, \eta((MN-1)T_{\mathrm{f}})]^T$ with the correlation matrix $\mathbb{E}\left[\boldsymbol{\eta}\boldsymbol{\eta}^H\right]=\sigma_0^2 \mathbf{G}$~\cite{ishihara2019svd,ishihara2021eigendecomposition}.
Moreover, $\mathbf{G} \in \mathbb{R}^{MN \times MN}$ is a Toeplitz matrix, whose first row and first column are denoted by $[g(0), g(-T_{\mathrm{f}}) \cdots, g(-(MN-1)T_{\mathrm{f}})]$ and $[g(0), g(T_{\mathrm{f}}) \cdots, g((MN-1)T_{\mathrm{f}})]^T$, respectively.
Furthermore, $\mathbf{H} \in \mathbb{C}^{MN \times MN}$ denotes the effective channels, which take into account the effects of both the dispersive channel and of the FTN-induced ISI. More specifically, the $k$th-row and $m$th-column entry of $\mathbf{H}$ is given by
\begin{IEEEeqnarray}{rCL}
\!\mathbf{H}(k,m)\!=\!\sum_{p=0}^{P-1} \!h_p e^{j 2 \pi \frac{(k_p+\kappa_p)\left(k-l_p\right)}{M N}} g\!\left(k T_{\mathrm{f}}-\left(m+l_p\right) T_{\mathrm{f}}\right)\!. \ \ \
\end{IEEEeqnarray}

Then, following the $M$-point FFT and SFFT, the received samples $\mathbf{z}$ in the TF domain
$\mathbf{Z}=\mathrm{vec}^{-1}(\mathbf{z})\in \mathbb{C}^{M \times N}$ are converted into those in the DD domain $\mathbf{Y}=\mathbf{F}_M^{H}\left(\mathbf{F}_M \mathbf{Z}\right) \mathbf{F}_N \in \mathbb{C}^{M \times N}$.
Furthermore, $\mathbf{Y}$ can be vectorized as follows:
\begin{IEEEeqnarray}{rCL}
\mathbf{y}&=&\left(\mathbf{F}_N \otimes \mathbf{I}_M\right) \mathbf{z}\nonumber \\
&=&\left(\mathbf{F}_N \otimes \mathbf{I}_M\right) \mathbf{H}\left(\mathbf{F}_N^H \otimes \mathbf{I}_M\right) \mathbf{x}_\mathrm{p}+\left(\mathbf{F}_N \otimes \mathbf{I}_M\right) \boldsymbol{\eta} \nonumber \\
&=&\mathbf{H}_{\mathrm{eq}}\mathbf{x}_\mathrm{p}+\boldsymbol{\eta}_{\mathrm{eq}}, \label{eq:dd}
\end{IEEEeqnarray}
where $\mathbf{H}_{\mathrm{eq}}=\left(\mathbf{F}_N \otimes \mathbf{I}_M\right) \mathbf{H}\left(\mathbf{F}_N^H \otimes \mathbf{I}_M\right) \in \mathbb{C}^{MN \times MN}$ and $\boldsymbol{\eta}_{\mathrm{eq}} =\left(\mathbf{F}_N \otimes \mathbf{I}_M\right) \boldsymbol{\eta}\in \mathbb{C}^{MN}$ represent the equivalent channel and the noise in the DD domain.

\section{Power Allocation for the Proposed OTFS-FTN}
\label{powerallocate}

\subsection{Mutual Information}

The upper bound of differential entropy with respect to the received frame $\mathbf{y}$ and the correlated noise $\boldsymbol{\eta}_{\mathrm{eq}}$ are given by \cite{cover1999elements}
\begin{IEEEeqnarray}{rCL}
\label{eq:h_e}
h_{\mathrm{e}}(\mathbf{y})& \leq &\log _2\left((\pi e)^{MN}\left|\mathbb{E}\left[\mathbf{y y}^H\right]\right|_{\mathrm{det}}\right) ,\\
\label{eq:h_n}
h_{\mathrm{e}}(\boldsymbol{\eta}_\mathrm{eq})&=&\log _2\left((\pi e)^{MN}\left|\mathbb{E}\left[\boldsymbol{\eta}_\mathrm{eq} \boldsymbol{\eta}_\mathrm{eq}^H\right]\right|_{\mathrm{det}}\right),
\end{IEEEeqnarray}
where $h_{\mathrm{e}}(\cdot)$ denotes the differential entropy and $e$ is the base of the natural logarithm.
Observe from \eqref{eq:dd} that the covariance matrix of the received frame is represented by
\begin{IEEEeqnarray}{rCL}
\mathbb{E}\left[\mathbf{y y}^H\right] &=& \mathbb{E}\left[\left(\mathbf{H}_{\mathrm{eq}} \mathbf{x}_\mathrm{p}+\boldsymbol{\eta}_\mathrm{eq}\right)\left(\mathbf{H}_{\mathrm{eq}} \mathbf{x}_\mathrm{p}+\boldsymbol{\eta}_\mathrm{eq}\right)^H\right] \nonumber \\
\label{eq:E_yy}&=& \mathbf{H}_{\mathrm{eq}} \mathbb{E}\left[\mathbf{x}_\mathrm{p} \mathbf{x}_\mathrm{p}^H\right] \mathbf{H}_{\mathrm{eq}}^H+\mathbb{E}\left[\boldsymbol{\eta}_\mathrm{eq} \boldsymbol{\eta}_\mathrm{eq}^H\right],
\end{IEEEeqnarray}
where we have
\begin{IEEEeqnarray}{rCL}
\mathbb{E}\left[\boldsymbol{\eta}_\mathrm{eq} \boldsymbol{\eta}_\mathrm{eq}^H\right]&=&\left(\mathbf{F}_N \otimes \mathbf{I}_M\right)\mathbb{E}[{\boldsymbol{\eta \eta}^H}] \left(\mathbf{F}_N^H \otimes \mathbf{I}_M\right) \nonumber \\
\label{eq:Geq}&=&\sigma_0^2\mathbf{G}_\mathrm{eq},
\end{IEEEeqnarray}
while $\mathbf{G}_\mathrm{eq}=\left(\mathbf{F}_N \otimes \mathbf{I}_M\right)\mathbf{G}\left(\mathbf{F}_N^H \otimes \mathbf{I}_M\right) \in \mathbb{C}^{MN \times MN}$.
From \eqref{eq:h_e}--\eqref{eq:Geq},
 we arrive at the upper-bound of the MI between the received frame and the precoded symbols in the form of:
\begin{IEEEeqnarray}{rCL}
I(\mathbf{x}_\mathrm{p} ; \mathbf{y})
&=&h_{\mathrm{e}}(\mathbf{y})-h_{\mathrm{e}}(\boldsymbol{\eta}_\mathrm{eq})\nonumber \\
&\leq& \log _2 \frac{(\pi e)^{MN}\left|\mathbb{E}\left[\mathbf{y y}^H\right]\right|_{\mathrm{det}}}{(\pi e)^{MN}\left|\mathbb{E}\left[\boldsymbol{\eta}_\mathrm{eq}\boldsymbol{\eta}_\mathrm{eq}^H\right]\right|_{\mathrm{det}}} \nonumber \\
&=& \log _2 \frac{\left|\mathbf{H}_{\mathrm{eq}} \mathbb{E}\left[\mathbf{x}_\mathrm{p}\mathbf{x}_\mathrm{p}^H\right] \mathbf{H}_{\mathrm{eq}}^H+ \sigma_0^2\mathbf{G}_\mathrm{eq}\right|_{\mathrm{det}}}{\left|\sigma_0^2\mathbf{G}_\mathrm{eq}\right|_{\mathrm{det}}}\nonumber \\
&=& \label{eq:MI}\log _2\!\left|\mathbf{I}_{MN}\!+\!\frac{1}{\sigma_0^2} \mathbf{H}_{\mathrm{eq}} \!\mathbb{E}\left[\mathbf{x}_\mathrm{p}\mathbf{x}_\mathrm{p}^H\right]\! \mathbf{H}_{\mathrm{eq}}^H \mathbf{G}_\mathrm{eq}^{-1}\!\right|_{\mathrm{det}}\!,
\end{IEEEeqnarray}
where the EVD of $\mathbf{G}_\mathrm{eq}$ is given by
\begin{IEEEeqnarray}{rCL}
\mathbf{G}_\mathrm{eq}=\mathbf{V} \boldsymbol{\Lambda} \mathbf{V}^H,
\end{IEEEeqnarray}
{while $\mathbf{V} \in \mathbb{C}^{MN \times MN}$ is a unitary matrix and $\boldsymbol{\Lambda}=\operatorname{diag}\left[\lambda_0, \cdots, \lambda_{MN-1}\right] \in \mathbb{R}^{MN \times MN}$ is a diagonal matrix whose diagonal elements are set in descending order.}
Hence, \eqref{eq:MI} can be rewritten as
\begin{IEEEeqnarray}{rCL}
I(\mathbf{x}_\mathrm{p} ; \mathbf{y}) & \leq &\!\log _2\!\left|\mathbf{I}_{MN}\!+\!\frac{1}{\sigma_0^2} \mathbf{H}_{\mathrm{eq}}\mathbb{E}\left[\mathbf{x}_\mathrm{p}\mathbf{x}_\mathrm{p}^H\right] \mathbf{H}_{\mathrm{eq}}^H \mathbf{V} \boldsymbol{\Lambda}^{-1} \mathbf{V}^H\right|_{\mathrm{det}}\! \nonumber\\
& =&\!\log _2\!\left|\!\mathbf{I}_{MN}\!+\!\frac{1}{\sigma_0^2} \!\mathbb{E}\!\left[\mathbf{x}_\mathrm{p}\mathbf{x}_\mathrm{p}^H\right] \!\mathbf{H}_{\mathrm{eq}}^H \mathbf{V} \boldsymbol{\Lambda}^{-\frac{1}{2}} \boldsymbol{\Lambda}^{-\frac{1}{2}} \mathbf{V}^H \mathbf{H}_{\mathrm{eq}}\!\right|_{\mathrm{det}} \nonumber\\
\label{eq:MI2} & =&\!\log _2\!\left|\!\mathbf{I}_{MN}\!+\!\frac{1}{\sigma_0^2}\!\mathbb{E}\!\left[\mathbf{x}_\mathrm{p}\mathbf{x}_\mathrm{p}^H\right] \mathbf{B}^H \mathbf{B}\!\right|_{\mathrm{det}}\!,
\end{IEEEeqnarray}
where $\mathbf{B}=\boldsymbol{\Lambda}^{-\frac{1}{2}} \mathbf{V}^H \mathbf{H}_{\mathrm{eq}}$.

{Based on the EVD, we have
$\label{BB}\mathbf{B}^H \mathbf{B}=\mathbf{U} \mathbf{\Xi} \mathbf{U}^H$,
where $\mathbf{U} \in \mathbb{C}^{MN \times MN}$ is a unitary matrix and $\boldsymbol{\Xi}=\operatorname{diag}\left[\xi_0, \cdots, \xi_{MN-1}\right] \in \mathbb{R}^{MN \times MN}$ is a diagonal matrix, whose diagonal elements are in descending order.} Then, \eqref{eq:MI2} is further rewritten by
\begin{IEEEeqnarray}{rCL}
I(\mathbf{x}_\mathrm{p} ; \mathbf{y}) & \leq &\log _2\left|\mathbf{I}_{MN}+\frac{1}{\sigma_0^2} \mathbb{E}\left[\mathbf{x}_\mathrm{p}\mathbf{x}_\mathrm{p}^H\right] \mathbf{U} \boldsymbol{\Xi} \mathbf{U}^H\right|_{\mathrm{det}} \nonumber \\
& =&\log _2\left|\!\mathbf{I}_{MN}+\frac{1}{\sigma_0^2} \boldsymbol{\Xi}^{\frac{1}{2}} \mathbf{U}^H \mathbb{E}\left[\mathbf{x}_\mathrm{p}\mathbf{x}_\mathrm{p}^H\right] \mathbf{U} \mathbf{\Xi}^{\frac{1}{2}}\!\right|_{\mathrm{det}}\!\!\!\!\!, \label{eq:MI3}
\end{IEEEeqnarray}
where we have $\boldsymbol{\Xi}^{\frac{1}{2}} \mathbf{U}^H \mathbb{E}\left[\mathbf{x}_\mathrm{p}\mathbf{x}_\mathrm{p}^H\right] \mathbf{U} \mathbf{\Xi}^{\frac{1}{2}}=\sigma_{\mathrm{x}}^2 \left(\mathbf{P}^H \mathbf{U} \boldsymbol{\Xi}^{\frac{1}{2}}\right)^H \mathbf{P}^H \mathbf{U}\boldsymbol{\Xi}^{\frac{1}{2}}$, which
is a positive semi-definite Hermitian matrix.

{Let us assume $\mathbb{E}\left[\mathbf{x x}^H\right]=\sigma_{\mathrm{x}}^2 \mathbf{I}_{MN}$.} Then, according to Hadamard’s inequality~\cite{horn2012matrix}, the expression in \eqref{eq:MI3} is maximized, when $\boldsymbol{\Xi}^{\frac{1}{2}} \mathbf{U}^H \mathbb{E}\left[\mathbf{x}_\mathrm{p}\mathbf{x}_\mathrm{p}^H\right] \mathbf{U} \mathbf{\Xi}^{\frac{1}{2}}$ is a diagonal matrix, which satisfies
\begin{IEEEeqnarray}{rCL}
\mathbf{U}^H \mathbb{E}\left[\mathbf{x}_\mathrm{p}\mathbf{x}_\mathrm{p}^H\right]\mathbf{U}&=&\sigma_{\mathrm{x}}^2 \boldsymbol{\Gamma}\nonumber \\
\mathbf{P}\mathbb{E}\left[\mathbf{x}\mathbf{x}^H\right]\mathbf{P}^{H}&=&\sigma_{\mathrm{x}}^2 \mathbf{U} \boldsymbol{\Gamma} \mathbf{U}^H\nonumber \\
\label{eq:P}\mathbf{P}&=&\mathbf{U}\boldsymbol{\Gamma}^\frac{1}{2},
\end{IEEEeqnarray}
where $\boldsymbol{\Gamma}=\operatorname{diag}\left[\gamma_0, \cdots, \gamma_{MN-1}\right] \in \mathbb{R}^{MN \times MN}$ is a real-valued diagonal matrix.

When the precoding matrix $\mathbf{P}$ satisfies \eqref{eq:P}, the MI of \eqref{eq:MI3} is upper-bounded by
\begin{IEEEeqnarray}{rCL}
I(\mathbf{x}_\mathrm{p} ; \mathbf{y}) & \leq& \log _2\left|\mathbf{I}_{MN}+\frac{\sigma_{\mathrm{x}}^2}{\sigma_0^2} \boldsymbol{\Xi}^{\frac{1}{2}} \boldsymbol{\Gamma} \boldsymbol{\Xi}^{\frac{1}{2}}\right|_{\mathrm{det}} \nonumber \\
\label{MImax}& =&\sum_{n=0}^{MN-1} \log _2\left(1+\frac{\sigma_{\mathrm{x}}^2 \gamma_n \xi_n}{\sigma_0^2} \right).
\end{IEEEeqnarray}

The transmitted energy of each frame in the proposed OTFS-FTN scheme is calculated by
\begin{IEEEeqnarray}{rCL}
E_N & =&\mathbb{E}\left[\int_{-\infty}^{\infty}\left|s(t)\right|^2 d t\right] \nonumber \\
& =&\mathbb{E}\left[\sum_k \sum_m s_k s_m^* g((k-m) T_\mathrm{f})\right] \nonumber \\
& =&\sigma_{\mathrm{x}}^2 \operatorname{tr}\left\{\boldsymbol{\Gamma} \mathbf{U}^H (\mathbf{F}_N \otimes \mathbf{I}_M) \mathbf{G}(\mathbf{F}_N^H \otimes \mathbf{I}_M)\mathbf{U}\right\}\nonumber \\
&=&\sigma_{\mathrm{x}}^2 \operatorname{tr}\{\boldsymbol{\Gamma} \boldsymbol{\Phi}\}\nonumber \\
\label{ED}&=& \sigma_{\mathrm{x}}^2 \sum_{n=0}^{MN-1} \gamma_n \phi_n,
\end{IEEEeqnarray}
where
\begin{IEEEeqnarray}{rCL}
\boldsymbol{\Phi}=\mathbf{U}^H (\mathbf{F}_N \otimes \mathbf{I}_M) \mathbf{G}(\mathbf{F}_N^H \otimes \mathbf{I}_M)\mathbf{U} \in \mathbb{C}^{MN \times MN},
\end{IEEEeqnarray}
and $\phi_n$$(n=0,\cdots, MN-1)$ denotes the $n$th diagonal entry of $\boldsymbol{\Phi}$.
Furthermore, ${\operatorname{tr}(\cdot)}$ denotes the trace operation.
Since the total power of the transmitted symbols is given by $E_N=MN \sigma_{\mathrm{x}}^2$, the expression in \eqref{ED} has to satisfy the following condition:
\begin{IEEEeqnarray}{rCL}
\sum_{n=0}^{MN-1} \gamma_n \phi_n=MN. \label{ED_c}
\end{IEEEeqnarray}

\subsection{Power Allocation}
To maximize the MI of \eqref{MImax} under the energy constraint of \eqref{ED_c}, the Lagrange function is given by
\begin{IEEEeqnarray}{rCL}
\label{Lfunction}J(\gamma_n,\lambda)\!&=&\!\sum_{n=0}^{MN-1} \!\log _2\!\left(\!1\!+\!\frac{\sigma_{\mathrm{x}}^2 \gamma_n \xi_n}{\sigma_0^2} \!\right)\!-\!\lambda\!\left(\!\sum_{n=0}^{MN-1}\! \gamma_n \phi_n\!-\!MN\!\right), \nonumber\\
\end{IEEEeqnarray}
where $\lambda$ denotes the Lagrange multiplier. To optimize the coefficients $\gamma_n$ $(n=0,\cdots,MN-1)$ in \eqref{Lfunction}, the following conditions have to be satisfied:
\begin{IEEEeqnarray}{rCL}
\frac{J(\gamma_n,\lambda)}{\partial \gamma_n}=0, \text { subject to } \gamma_n \geq 0.
\end{IEEEeqnarray}

Hence, the optimal coefficients $\gamma_n$ are represented by
\begin{IEEEeqnarray}{rCL}
\label{r_n}\gamma_n=\max \left(\frac{1}{(\ln 2) \lambda \phi_n }-\frac{\sigma_0^2}{\xi_n \sigma_{\mathrm{x}}^2}, 0\right),
\end{IEEEeqnarray}
which can be solved by the classic water-filling algorithm similar to that used in the singular-value decomposition (SVD)-based multiple-input multiple-output (MIMO) systems \cite{goldsmith2005wireless}.

Upon normalizing \eqref{MImax} by the frame duration $MN\alpha T_0$ and the associated bandwidth $2W(1+\beta)$, the information rate of EVD-precoded OTFS-FTN signaling is given by
\begin{IEEEeqnarray}{rCL}
R\!=\!\frac{1}{2W(1+\beta)MN \alpha T_0 } \!\sum_{n=0}^{MN-1} \!\log _2\left(\!1\!+\!\frac{\sigma_{\mathrm{x}}^2 \gamma_n \xi_n}{\sigma_0^2}\!\right)\left[\mathrm{bps}/\mathrm{Hz}\right], \nonumber\\
\end{IEEEeqnarray}
which corresponds to the achievable bound. 
\subsection{Demodulation}
Upon multiplying $\mathbf{y}$ by the weight matrix $\mathbf{D}=\mathbf{U}^H \mathbf{B}^H \boldsymbol{\Lambda}^{-\frac{1}{2}} \mathbf{V}^H \in \mathbb{C}^{MN \times MN}$, we arrive at
\begin{IEEEeqnarray}{rCL}
\mathbf{y}_{\mathrm{d}} & =&\mathbf{D} \mathbf{y} \nonumber \\
& =&\mathbf{U}^H \mathbf{B}^H \boldsymbol{\Lambda}^{-\frac{1}{2}} \mathbf{V}^H \mathbf{H}_{\mathrm{eq}} \mathbf{x}_{\mathrm{eq}}+\mathbf{D} \boldsymbol{\eta}_{\mathrm{eq}}\nonumber \\
\label{yd}& =&\boldsymbol{\Xi} \boldsymbol{\Gamma}^{\frac{1}{2}} \mathbf{x}+\boldsymbol{\eta}_{\mathrm{d}},
\end{IEEEeqnarray}
where we have $\boldsymbol{\eta}_{\mathrm{d}}=\mathbf{D} \boldsymbol{\eta}_{\mathrm{eq}} \in \mathbb{C}^{MN}$, associated with a diagonal correlation matrix, yielding:
\begin{IEEEeqnarray}{rCL}
\mathbb{E}\left[\boldsymbol{\eta}_{\mathrm{d}} \boldsymbol{\eta}_{\mathrm{d}}^H\right] & =&\mathbf{D} \mathbb{E}\left[\boldsymbol{\eta}_{\mathrm{eq}} \boldsymbol{\eta}_{\mathrm{eq}}^H\right] \mathbf{D}^H \nonumber \\
& =&\sigma_0^2 \mathbf{U}^H \mathbf{B}^H \boldsymbol{\Lambda}^{-\frac{1}{2}} \mathbf{V}^H \mathbf{G}_{\mathrm{eq}} \mathbf{V} \boldsymbol{\Lambda}^{-\frac{1}{2}} \mathbf{B U} \nonumber \\
\label{nd}& =&\sigma_0^2 \mathbf{\Xi}.
\end{IEEEeqnarray}

Observe from \eqref{yd} and \eqref{nd} that the effective channel is diagonalized, and the correlated noise is whitened. Therefore, efficient symbol-by-symbol demodulation can be carried out.
More specifically, let us consider the $n$th received symbol $y_{\mathrm{d},n}$ in \eqref{yd}. Then, the probability density function for calculating the corresponding log-likelihood ratio (LLR) is given by
\begin{IEEEeqnarray}{rCL}
p\left(y_{\mathrm{d}, n} \mid x_n\right)=\exp \left(-\frac{\left|y_{\mathrm{d}, n}-\xi_n \sqrt{\gamma_n} x_n\right|^2}{\xi_n \sigma_0^2}\right).
\end{IEEEeqnarray}
\section{Performance Results}
\label{simulation}

\begin{figure}
\centering
\includegraphics[width=.7\linewidth]{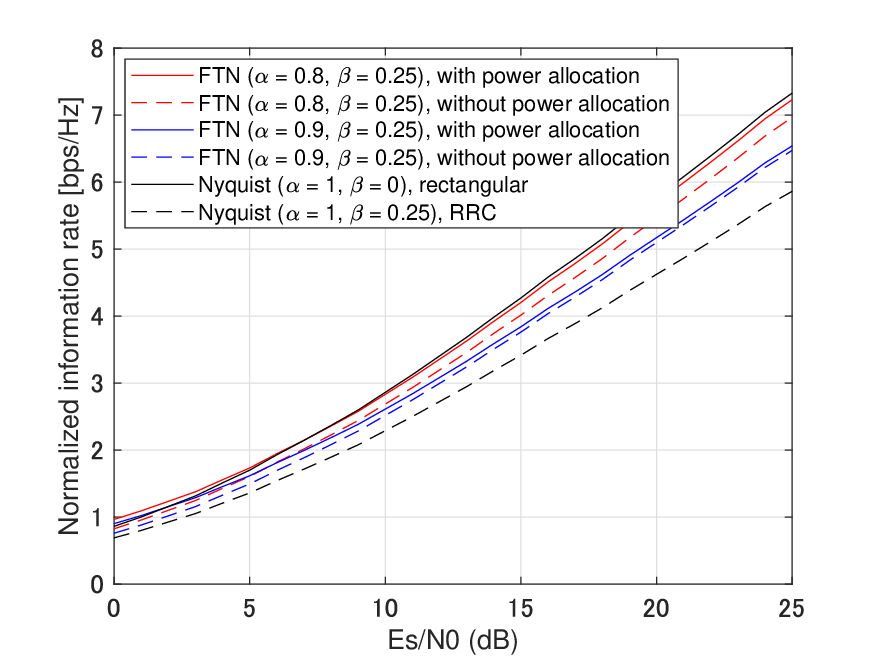}
\caption{Normalized information rate of the proposed OTFS-FTN signaling scheme with and without PA. The information rates of the conventional Nyquist-based OTFS schemes with the ideal rectangular pulse filter ($\beta=0$) and the RRC filter ($\beta=0.25$) are also plotted.}
\label{Inf}
\end{figure}
Fig. \ref{Inf} shows the normalized information rate of the proposed OTFS-FTN signaling scheme with and without PA, each employing the RRC shaping filter having a roll-off factor of $\beta=0.25$. {Also, we limit the range of packing ratio to $\alpha\ge1/(1+\beta)$ for avoiding an ill-conditioned case~\cite{ishihara2021evolution}.} {The basic system parameters are set to $(P, M, N)=(20,128,12)$, and the maximum integer Doppler-shift tap is given by $k_\mathrm{max}=5 \ (\ge k_p)$. The Nyquist-based symbol interval is normalized to $T_0=1$ for simplicity.}
In the proposed scheme operating without PA, the precoding matrix is set to $\boldsymbol{\Gamma}=\mathbf{I}_{MN}$, i.e., $\mathbf{P}=\mathbf{U}$.
The symbol packing ratio was set to $\alpha=0.9$ and $0.8$.
Moreover, the conventional Nyquist-based OTFS signaling scheme {($\alpha=1$)} employing the same RRC filter having $\beta=0.25$ is chosen as a benchmark. {Furthermore, the upper bound employing the ideal rectangular shaping filter ($\beta=0$) is also considered.}
Observe in Fig. \ref{Inf} that the proposed OTFS-FTN signaling scheme relying on PA outperformed the {Nyquist-based} scheme employing {the} RRC filter, as well as the proposed scheme dispensing with PA while approaching the ideal upper bound associated with the rectangular filter $(\beta=0)$.
\begin{figure}[tbp]
\subfigure[]{
\centering
\includegraphics[width=0.46\linewidth]{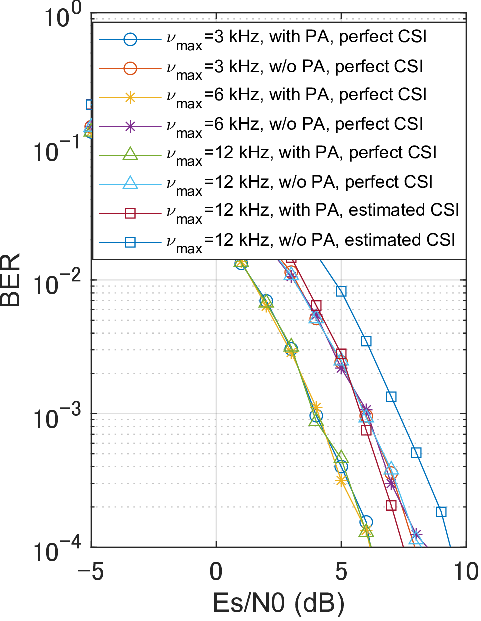}
}
\subfigure[]{
\centering
\includegraphics[width=0.38\linewidth]{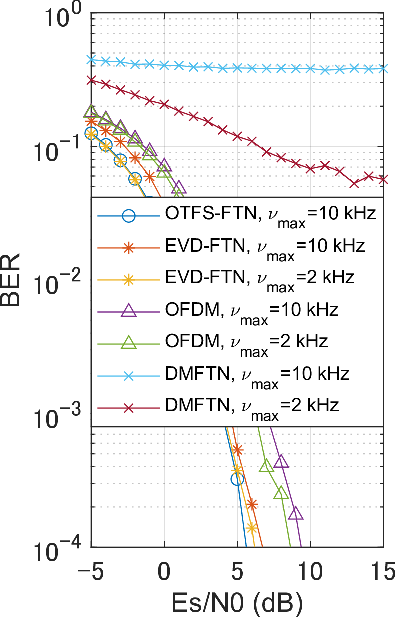}
}
\caption{{BER performance; (a) perfect/estimated CSI, (b) comparisons with EVD-FTN signaling \cite{ishihara2021eigendecomposition}, DMFTN signaling \cite{ishihara2023differential}, and OFDM.}}
\label{BER01}
\end{figure}
\begin{figure}[tbp]
\subfigure[]{
\centering
\includegraphics[width=0.4\linewidth]{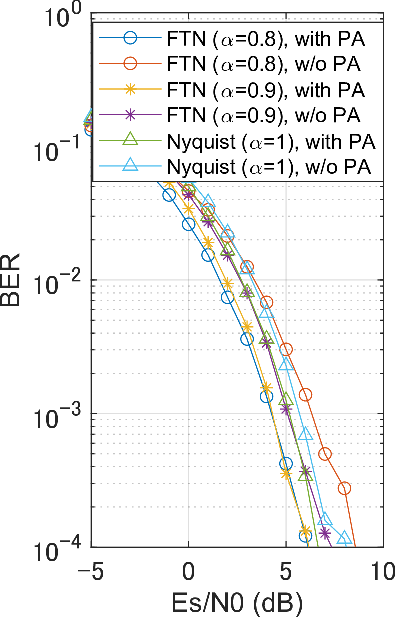}
}
\hspace{0.01\textwidth}
\subfigure[]{
\centering
\includegraphics[width=0.4\linewidth]{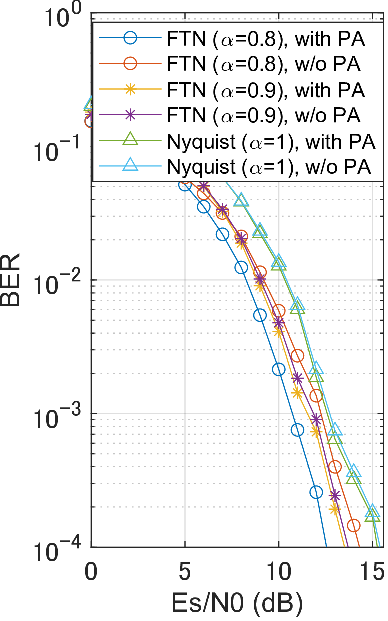}
}
\caption{{BER performance for different packing ratios with the fixed transmission rate. (a) $R_\mathrm{t}=1.5$ bps/Hz. (b) $R_\mathrm{t}=3$ bps/Hz.}}
\label{BERtrans}
\end{figure}

{Next, to evaluate the achievable BER performance, we considered the extended vehicular A (EVA) model \cite{3gpp.36.101} with the number of channel taps $P=9$. Each channel tap has a single Doppler shift generated through Jakes' formula $\nu_p=\nu_{\max} \cos \left(\theta_p\right)$, where $\theta_p$ is uniformly distributed over $[-\pi, \pi]$.
Furthermore, to achieve a near-capacity BER performance, a 3/4-rate low-density parity check (LDPC) coding scheme associated with a maximum of 50 iterations was used for every transmission frame.

{Fig.~\ref{BER01}(a) shows the BERs of the proposed OTFS-FTN signaling scheme for the different maximum Doppler shifts of $\nu_{\rm max}= 3$, $6$, and $12$ kHz. Furthermore, we considered QPSK, $\Delta f=30$ kHz, and $(M, N,\alpha,\beta)=(64,30,0.82,0.25)$.}
{In addition to the perfect channel state information (CSI) scenario, we also considered the scenario where DD-domain channel estimation of \cite{10453468} is employed at the transmitter under the assumption of channel reciprocity.
Observe in Fig.~\ref{BER01}(a) that the proposed scheme associated with PA outperformed that without PA even
in the presence of channel estimation errors, while achieving a performance close to that of the scheme without PA and with perfect CSI.}
{Fig.~\ref{BER01}(b) compares the proposed scheme to the three benchmarks, i.e., the EVD-based FTN (EVD-FTN) signaling \cite{ishihara2021eigendecomposition}, open-loop differential multi-carrier FTN (DMFTN) signaling \cite{ishihara2023differential}, and conventional OFDM. We considered QPSK, $\Delta f=60$ kHz, and $(M, N,\alpha,\beta)=(256,6,0.85,0.25)$. It can be seen in Fig.~\ref{BER01}(b) that the proposed scheme outperformed the benchmarks in the simulated scenario.}

{Fig.~\ref{BERtrans}(a) and (b) show the BERs of the proposed OTFS-FTN scheme under different packing ratios while employing the same RRC filter ($\beta=0.25$).} {The parameters are set as $\Delta f=30$ kHz, $\nu_{\max}=7.5$ kHz and $(M, N)=(128,12)$.} The transmission rate is given by
$R_{\mathrm{t}}=\frac{3}{4} \cdot \frac{1}{2 W(1+\beta)} \cdot \frac{1}{MN \alpha T_0} \cdot \sum_{n=0}^{MN-1} b_n$ [$\mathrm{bps} / \mathrm{Hz}$],
where $b_n$ denotes the number of bits mapped onto the $n$th symbol, and the coefficient 3/4 represents the coding rate.
The transmission rates of $R_\mathrm{t}=1.5$ and $3$ bps/Hz are considered.
{Based on the bit-loading concept of \cite{ishihara2019svd,ishihara2021eigendecomposition} and on \eqref{r_n}, either QPSK, 16–QAM, 64–QAM, or 256–QAM is assigned to each activated symbol to achieve the target transmission rate $R_\mathrm{t}$.}
{It can be seen from Fig. \ref{BERtrans}(a) that the proposed scheme using PA ($\alpha=0.8$ and $0.9$) outperformed the Nyquist-based benchmark. Furthermore, in Fig. \ref{BERtrans}(b), the reduction in the packing ratio improved the BER performance at $R_\mathrm{t}=3$~bps/Hz.}
\section{Conclusions}
\label{conclusion}
In this paper, we proposed a new EVD-precoded OTFS modulation scheme in the context of FTN transmission under the doubly selective fading channels.
Based on EVD-aided diagonalization, efficient symbol-by-symbol demodulation is achieved in the DD domain.
Furthermore, by maximizing the mutual information of the proposed OTFS-FTN scheme, we derived the PA coefficients, maximizing MI.
Our performance results demonstrated that the proposed FTN-OTFS scheme exhibited a higher information rate than its conventional Nyquist-based time-orthogonal signaling counterpart using the same RRC shaping filter. 
%

\bibliographystyle{IEEEtran}
\bibliography{IEEEabrv,ref}

\end{document}